\documentclass[aps,reprint, prl, superscriptaddress, longbibliography,showpacs,nobibnotes,floatfix]{revtex4-1}
\usepackage{graphicx,xcolor}% Include figure files [dvips]
\usepackage{dcolumn}% Align table columns on decimal point
\usepackage{bm}% bold math
\usepackage{graphics}
\usepackage{float}
\usepackage{epsfig}
\usepackage{multirow}
\usepackage{amsmath}
\usepackage{multirow}
\usepackage{tabularx}
\usepackage{bm}        % for math
\usepackage{mathrsfs}
\usepackage{tabularx}
\usepackage{tabulary}
\usepackage{gensymb}
\usepackage{microtype}
\usepackage{amssymb}
\usepackage{bm}
\usepackage{braket}
\usepackage{graphicx}
\usepackage[dvipsnames]{xcolor}
\usepackage[pass,paperwidth=8.5in,paperheight=11in]{geometry}
\usepackage[linktocpage=true]{hyperref}
\usepackage{hyperref}
\usepackage{comment}
\usepackage{soul}
\usepackage{notes2bib}
\hypersetup{
  pdfnewwindow=true, 
  colorlinks=true,
  linkcolor=blue, 
  anchorcolor=blue,
  citecolor=blue, 
  filecolor=blue,
  menucolor=blue, 
  urlcolor=blue}

\usepackage[normalem]{ulem}
\usepackage{xcolor}
\usepackage{orcidlink}

\newcommand{\vk}{\mathbf{k}}
\newcommand{\vG}{\mathbf{G}}
\newcommand{\vb}{\mathbf{b}}

\newcommand{\vsig}{\bm{\sigma}}
\newcommand{\hv}{\hbar v}
\newcommand{\Ztwo}{Z_2}

\begin{document}

\title{Relaxation-driven topological domains in moir\'e materials}

% \title{Programmable topological domains and edge states from relaxation-driven patterning in \\moir\'e  materials}

%\title{Programmable topological domains and edge states via relaxation-driven \\moir\'e patterning}

%\title{Structural relaxation drives programmable topological domains and edge states in \\moir\'e materials}

%\title{Interlayer relaxation dynamics and emergent topological domain formation in \\moir\'e superlattices}
%\title{Dynamical Topological Phase Transitions in moir\'e Superlattices}

\author{Arjyama Bordoloi\,\orcidlink{0009-0006-2760-3866}}
\affiliation{Department of Mechanical Engineering, University of Rochester, Rochester, New York 14627, USA}

\author{Daniel Kaplan\,\orcidlink{0000-0002-3957-0030}}
\email{d.kaplan1@rutgers.edu}
\affiliation{Center for Materials Theory, Department of Physics and Astronomy, Rutgers University, Piscataway, New Jersey 08854-8019, USA}

\author{Sobhit Singh\,\orcidlink{0000-0002-5292-4235}}
\email{s.singh@rochester.edu}
\affiliation{Department of Mechanical Engineering, University of Rochester, Rochester, New York 14627, USA}
\affiliation{Materials Science Program, University of Rochester, Rochester, New York 14627, USA}
%\affiliation{Center for Coherence and Quantum Optics, University of Rochester, Rochester, NY, 14627, USA}

\date{\today}

\begin{abstract}

Spatial control of topology is highly desirable for realizing tunable quantum functionalities in materials. Moir\'e superlattices formed by twisting van der Waals heterostructures provide a natural platform for spatially modulated electronic phases, yet the emergence of tunable topological domains in these systems remains largely unexplored. Here we show that structural relaxation in twisted bilayer BiSb drives the formation of a distinct moiré topological phase, 
characterized by coexisting topologically nontrivial (Z$_2$ = 1) and trivial (Z$_2$ = 0) domains within a single moiré unit cell. As the twist angle is reduced, relaxation-induced modulation of the interlayer separation stabilizes an expanding network of topological regions embedded within trivial backgrounds of the moiré unit cell. 
The resulting internal domain boundaries host topologically-protected gapless 1D edge states that are directly visible in our simulated scanning-tunnelling microscopy maps.Furthermore, we demonstrate that the real-space topological domain structure and associated gapless edge states can be reversibly tuned by an out-of-plane electric field. Together, these results establish twisted BiSb as a promising platform for programmable topological domain patterning, where intrinsic networks of edge channels can be continuously tuned and electrically reconfigured in moir\'e materials.

\end{abstract}

\maketitle

\begin{figure*}[hbtp]
\centering
\includegraphics [width=0.95\textwidth]{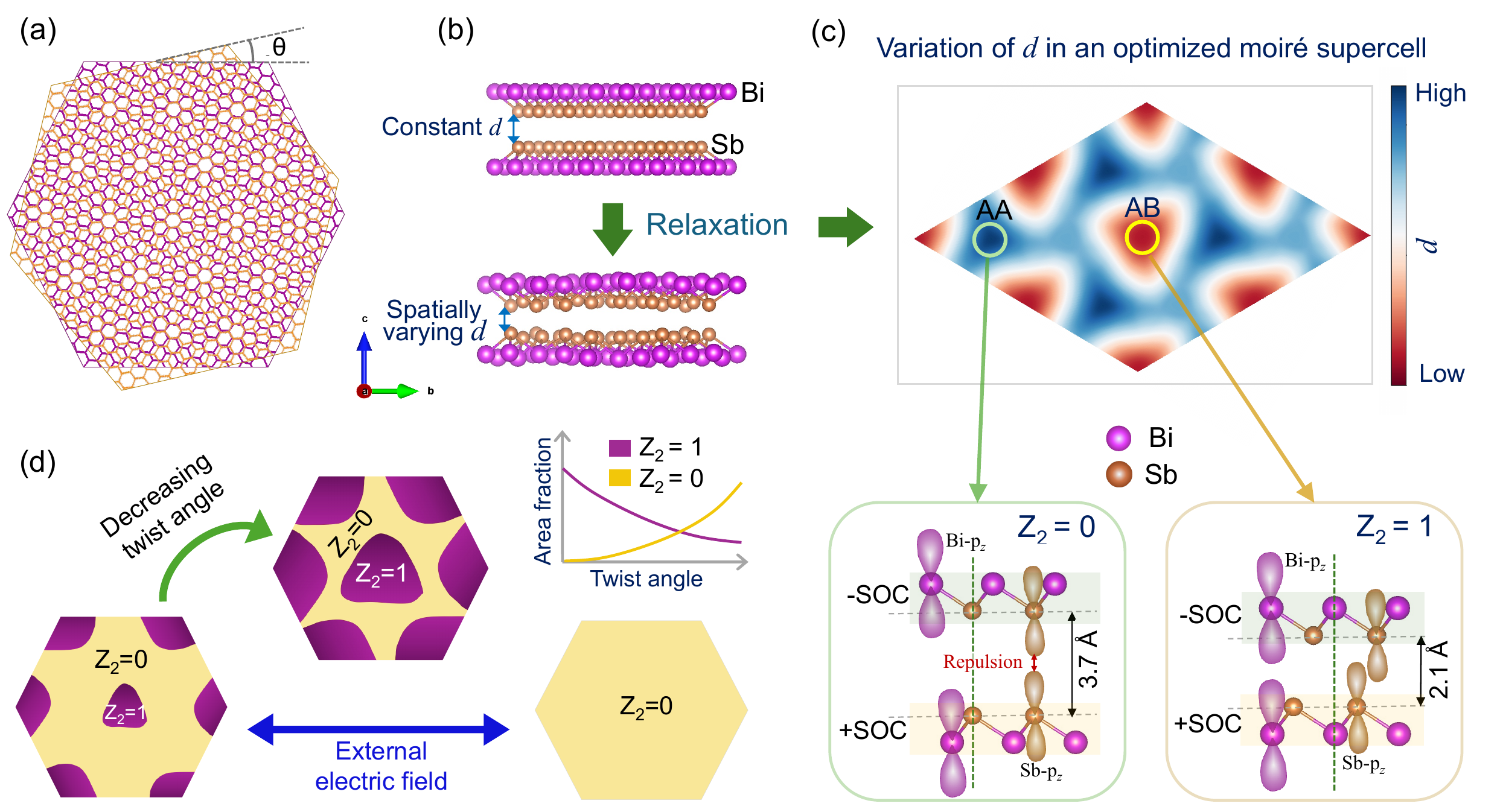}
\caption{(a)\,Schematic of the moir\'e supercell formed by twisting two BiSb layers by an angle $\theta$. (b)\,BiSb–SbBi bilayer at a 13.17$\degree$ twist angle, illustrating how structural relaxation leads to spatially varying interlayer separation\,\textit{d}. (c)\,2D phase map of \textit{d} across the moir\'e supercell, where red and blue indicate regions of lower and higher interlayer separation, respectively. The minimum and maximum values of $d$ correspond to high-symmetry stacking configurations AB (Bi atoms in the two layers aligned) and AA (Sb atoms in the two layers aligned), respectively. In the AA region, enhanced Sb-p$_z$ orbital repulsion leads to an increased interlayer spacing. (d) Schematic illustrating the role of twist angle and an out-of-plane electric field in controlling topological domain size: decreasing the twist angle enlarges the Z$_2$\,=\,1 domains, while an applied electric field can drive a transition to a fully Z$_2$\,=\,0 state.}
\label{fig:Figure1}
\end{figure*}

\noindent \textbf{Introduction.} 
Twisted two-dimensional van der Waals (vdW) heterostructures offer a highly versatile and tunable platform for exploring a wide range of emergent quantum phenomena, including strongly correlated insulating phases~\cite{Shimazaki_Nature_2020, Wang_Nature_mat_2020}, the fractional anomalous quantum Hall effect~\cite{Park_Nature_2023, Cai_Nature_2023}, superconductivity~\cite{Xia_Nature_2025, Guo_Nature_2025}, and generalized Wigner crystals~\cite{Zhou_Nature_2021, Li_Nature_2021, Regan_Nature_2020}. When two layers are twisted by an angle $\theta$, the interference of their lattice periodicities produces a moir\'e superlattice characterized by spatially varying atomic registries~\cite{Duncan_Nature_2025, Maity_PRB_2021}. 
These local variations induce periodic modulations of interlayer hybridization and stacking energies across the supercell, producing an energy landscape that is highly sensitive to local atomic configuration. Such variations strongly renormalize the electronic structure, leading to domain-dependent properties that can be precisely tuned by adjusting the twist angle. Thus, twisted 2D vdW heterostructures provide an exciting and highly controllable platform for realizing materials in which contrasting quantum phases can coexist within the same system maintaing their key features~\cite{Weston_Nature_Nanotech_2022, Rosenberger_ACS_nano_2020, Enaldiev_2DMaterials_2021}.
%Consequently, a moir\'e heterostructure is not just a rigid stacking of layers but undergoes local atomic relaxation within each domain~\cite{Duncan_Nature_2025, Maity_PRB_2021}. 

%An intriguing consequence is the possibility of realizing spatially modulated topological phases. 
In this context, a promising direction is the design of materials with spatially varying local topological character. 
Unlike conventional 2D topological insulators, where edge states are confined only to sample boundaries, moir\'e systems can host internal topological interfaces arising purely from local variations in stacking. This gives rise to self-organized networks of conducting 1D edge states along domain boundaries; 
these states are analogous to lithographically-patterned topological circuitry, but are instead intrinsically defined by the moir\'e geometry, ensuring high symmetry and uniformity. 
Previously, these topological mosaic networks were found to be especially important in the description of Landau levels~\cite{chalker1988percolation}.
Without a magnetic field, such in-plane topological networks provide a natural platform for planar device architectures, where conducting 1D channels can be electrically tuned and reconfigured within the same material. 
%while maintaining conducting and dispersing states.

Although topological mosaic phases have been proposed theoretically in twisted materials, existing studies~\cite{Tong_Nature_Phy_2017, Tateishi_Phys_Rev_Research_2022,SanJose2013,Efimkin2018} have largely neglected the role of structural relaxation, despite its central importance in determining the local electronic structure of moir\'e heterostructures~\cite{Carr_NaturE_Rev_Materials_2020, Carr_Phy_Rev_B_2018, Nam_PRB_2020,koshino2020effective,nakatsuji2023multiscale}.
Structural relaxation fundamentally reconstructs moiré superlattices by expanding energetically favorable stacking regions and compressing unfavorable registries into narrow domain walls~\cite{Nam_PRB_2020, Shi_PRL_2026}. 
This process induces pronounced spatial variations in interlayer separation and electronic hybridization, leading to an emergent real-space topological patterning in moiré materials.

Here we show that structural relaxation can itself drive the emergence of real-space topological domain patterning in twisted bilayer BiSb heterostructures. 
Although an isolated BiSb monolayer is topologically trivial~\cite{Singh_PRB_2017,kaplan_2024_arxiv, Bordoloi_2DMaterials_2025}, twisting inverted BiSb layers generates a moiré superlattice in which relaxation induces substantial modulation of the interlayer spacing and hybridization (see Fig.~\ref{fig:Figure1}). 
This gives rise to a rich 
%landscape of programmable 
network of topologically trivial (Z$_2$\,=\,0) and nontrivial (Z$_2$\,=\,1)  domains within a single moir\'e supercell. These topological domains persist even at relatively large twist angles (e.g., 21.78$\degree$). 
%The relative area fraction of Z$_2$\,=\,1 regions increases as the twist angle decreases. 
The resulting domain boundaries host robust, gapless edge states that are directly visible in simulated scanning-tunnelling microscopy (STM) maps through enhanced local density of states (LDOS). 
We further demonstrate that the area of the topological domains can be continuously tuned by twist angle, while an out-of-plane electric field enables reversible control of the edge-state network. 
%Our results establish relaxation-driven moiré reconstruction as a general mechanism for engineering reconfigurable topological phases and intrinsic edge-channel architectures in moir\'e materials.
Overall, our results establish that topology in moiré systems is not {\it globally} uniform, instead it emerges as a real-space texture governed by the interplay of twist, lattice relaxation, and spin-orbit coupling.  \\

\begin{figure*}[hbtp]
\centering
\includegraphics [width=1\textwidth]{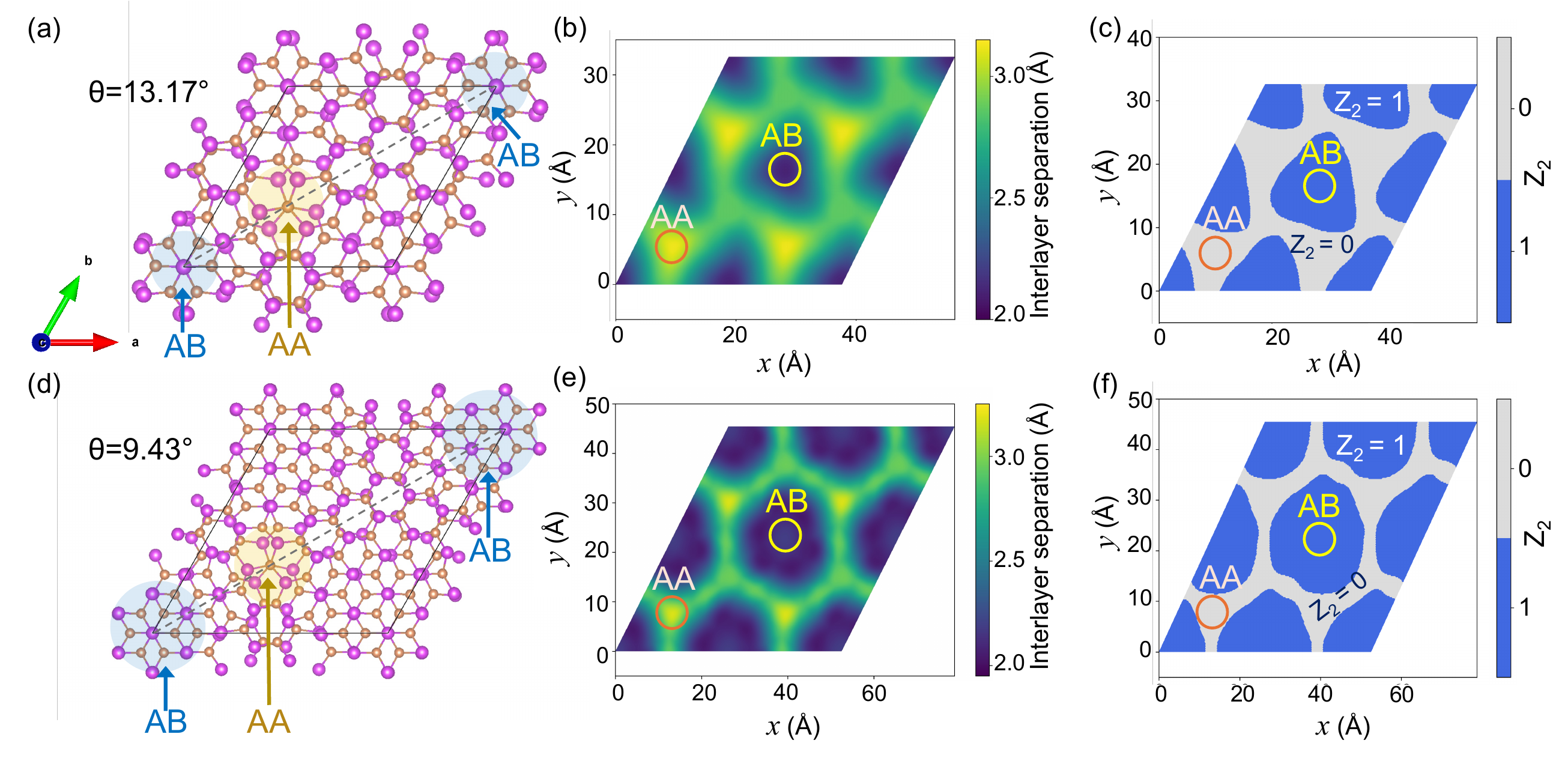}
\caption{Optimized crystal structures of twisted BiSb–SbBi bilayers at (a) 13.17$\degree$ and (d) 9.43$\degree$ twist angles, with AB and AA stacking regions highlighted in blue and yellow, respectively. 2D phase maps of interlayer separation (\textit{d}) for 2\,$\times$2\,$\times$1 supercells at (b) 13.17$\degree$ and (e) 9.43$\degree$ twists. 2D maps showing the spatial variation of the Z$_2$ invariant for the bilayers at (c) 13.17$\degree$ and (f) 9.43$\degree$ twist angles.}
\label{fig:phase diagram}
\end{figure*}

\noindent \textbf{Results}

\noindent\textbf{Topological properties of  twisted BiSb-SbBi bilayers.}
The BiSb monolayer crystallizes in the $p3m1$ symmetry (layer group no.\,69). Due to the lack of inversion symmetry, it exhibits strong Rashba spin–orbit coupling (RSOC) ($\alpha_R$\,=\,1.9\,eVÅ)~\cite{Bordoloi_2DMaterials_2025}. When two BiSb monolayers are stacked in an inverted configuration, the resulting BiSb–SbBi bilayer adopts $p\bar{3}m1$ symmetry (layer group no.\,72). In this bilayer, inversion symmetry is restored, giving rise to a hidden Rashba effect~\cite{Riley_Nature_Phy_2014, Yuan_Nat_comm_2019}, where RSOC exists locally within each layer but cancels out globally due to inversion symmetry. While the monolayer is topologically trivial (Z$_2$\,=\,0), the untwisted bilayer is topologically nontrivial (Z$_2$\,=\,1)~\cite{Bordoloi_2DMaterials_2025}.

Twisted bilayer configurations were generated by rotating one BiSb monolayer relative to the other at specific twist angles. In this work, we explore three commensurate angles: 21.78$\degree$, 13.17$\degree$, and 9.43$\degree$, chosen based on the commensurate moir\'e condition, 
$cos\theta=\frac{m^2 + 4mn + n^2}{2(m^2 + mn + n^2)},$ 
where $m$ and $n$ are positive integers. The results for the 13.17$\degree$ and 9.43$\degree$ twist angles are presented in the main text, while the 21.78$\degree$ case is included in the Supplementary Information. For all three twist angles, the relaxed moir\'e supercells crystallize in the $p312$ symmetry (layer group no.\,67), containing 28, 76, and 128 atoms for the 21.78$\degree$, 13.17$\degree$, and 9.43$\degree$ twist angles, respectively. While inversion symmetry is broken in the moir\'e supercell, the threefold rotational symmetry is preserved.\\

\noindent\textbf{Role of structural relaxation in determining the local topological character of  twisted BiSb-SbBi bilayers.} 
During structural relaxation, the twisted bilayers undergo significant lattice reconstruction, resulting in spatially modulated atomic registries across the moir\'e supercell. Along the long diagonal of the supercell, two high-symmetry stacking configurations are observed: AB, where Bi atoms from the two layers align, and AA, where Sb atoms from both layers lie directly on top of each other, as shown in Figure~\ref{fig:phase diagram}(a) and ~\ref{fig:phase diagram}(d). The regions separating these two domains exhibit intermediate stacking configurations that smoothly interpolate between the two high-symmetry configurations. As shown in Figures.~\ref{fig:phase diagram}(b) and ~\ref{fig:phase diagram}(e), the optimized structures exhibit a continuous modulation of the interlayer separation \textit{d}, which directly determines the local topological character, as reflected in the Z$_2$ phase diagrams [Figs.~\ref{fig:phase diagram}(c) and ~\ref{fig:phase diagram}(f)]. Notably, at larger twist angles such as 60$\degree$, which correspond to Bernal stacking, the optimized bilayer remains topologically trivial and does not exhibit any spatially varying topological character (see Supplementary Fig.~S2). This is a concrete example of a topological phase transition that is purely driven by twisting.

To elucidate the spatial variation of \textit{d} across the moir\'e supercell and its influence on the topological properties, we analyze stacking configurations of the untwisted bilayer, including the AA and AB arrangements based on sampling the disregistry index of the layers \cite{carr2017twistronics,Carr_Phy_Rev_B_2018,cazeaux2023relaxation} by constructing a 15\,$\times$\,15 grid generated through systematic relative shifts of the top layer with respect to the bottom layer along random in-plane directions, i.e., ($x$, $y$)\,$\rightarrow$\,($x+au_x$, $y+au_y$), where  $a$ represents the lattice constant. Among all configurations, AB stacking is the most energetically favorable, whereas AA stacking is the least favorable, lying approximately 0.5 eV higher in energy. The relative energies of all intermediate configurations are displayed as a 2D colormap in Supplementary Figure\,1(a).
%%%%%%%%

The interlayer separation between the two BiSb layers is highly sensitive to the local stacking configuration. In the AA stacking, the vertical alignment of Sb atoms leads to significant overlap of the out-of-plane Sb–$p_z$ orbitals [as shown in Figure~\ref{fig:Figure1}(d)], resulting in strong interlayer repulsion and, consequently, a large interlayer separation of approximately 3.7\,Å. In contrast, the AB stacking, where Sb atoms are laterally displaced, minimizes orbital overlap, reducing interlayer repulsion and yielding a smaller separation of about 2.1\,\AA\ ; see Supplementary Fig.\,1(b) for details. 

Owing to the strong Sb–$p_z$ orbital repulsion, the electronic states near the Fermi level are primarily contributed by the Bi–$p_z$ orbitals. Since the stacking configuration strongly influences the interlayer separation, it directly governs the degree of hybridization and interlayer electron tunneling between the Bi–p$_z$ orbitals. As demonstrated in our previous work~\cite{Bordoloi_2DMaterials_2025}, the interplay between this interlayer tunneling and the SOC strength determines the topological character of the system. Consequently, the AB stacking, which exhibits the smallest interlayer spacing, shows the strongest electron tunneling and exhibits a topologically nontrivial phase (Z$_2$\,=\,1). In contrast, AA stacking, characterized by the largest interlayer separation, exhibits reduced tunneling and remains topologically trivial (Z$_2$\,=\,0) in their optimized configuration.

These stacking-dependent variations in \textit{d} and Z$_2$ are directly reflected in the moir\'e supercells. In unrelaxed twisted bilayers, the interlayer separation is uniform throughout the supercell~[Figure~\ref{fig:Figure1}(b)]. However, structural relaxation induces spatial variations in the interlayer distance due to lattice reconstruction, resulting in locally distinct stacking configurations. For instance, the corners of the moir\'e supercell correspond to AB stacking (blue-shaded regions), while the yellow-shaded regions correspond to AA stacking. Since each stacking configuration exhibits a characteristic interlayer separation, so the relaxed supercell shows an in-plane modulation of interlayer separation. Figure~\ref{fig:phase diagram} illustrates this variation for moir\'e supercells with twist angles of 13.17$\degree$ and 9.43$\degree$. AB-stacked regions exhibit the smallest interlayer separation, whereas AA-stacked regions have the largest. Overall, the interlayer distance across the supercell ranges approximately from 2.1\,Å to 3.2\,Å.

%Spatial variations in interlayer separation influence interlayer hybridization and electron tunneling in the moir\'e supercell. 
Leveraging the previously established dependence of the Z$_2$ topological character on both stacking pattern and interlayer distance, we constructed a spatial Z$_2$ phase diagram for twist angles of 13.17$\degree$ and 9.43$\degree$, shown in Figs.~\ref{fig:phase diagram}(c) and ~\ref{fig:phase diagram}(f). As expected, AA stacking regions with the largest interlayer separation are topologically trivial (Z$_2$\,=\,0, shown in gray), whereas AB domains with the smallest interlayer separation exhibit a nontrivial phase (Z$_2$\,=\,1, blue regions).
%As expected, the spatial distribution of the Z$_2$ invariant mirrors the interlayer separation map. 

Notably, reducing the twist angle from  13.17$\degree$ to 9.43$\degree$ increases the Z$_2$\,=\,1 domain fraction from $\sim$\,49\% to approximately 70\% of the total area. This arises because, during structural relaxation, the 9.43$\degree$ twisted bilayer significantly maximizes the AB domains in the optimized structure, owing to the lower energy of the AB stacking configuration. This suggests that the size of the topological domains in BiSb-SbBi bilayers can be effectively controlled by adjusting the twist angle.\\

\begin{figure}[!t]
\centering
\includegraphics [width=1\columnwidth]{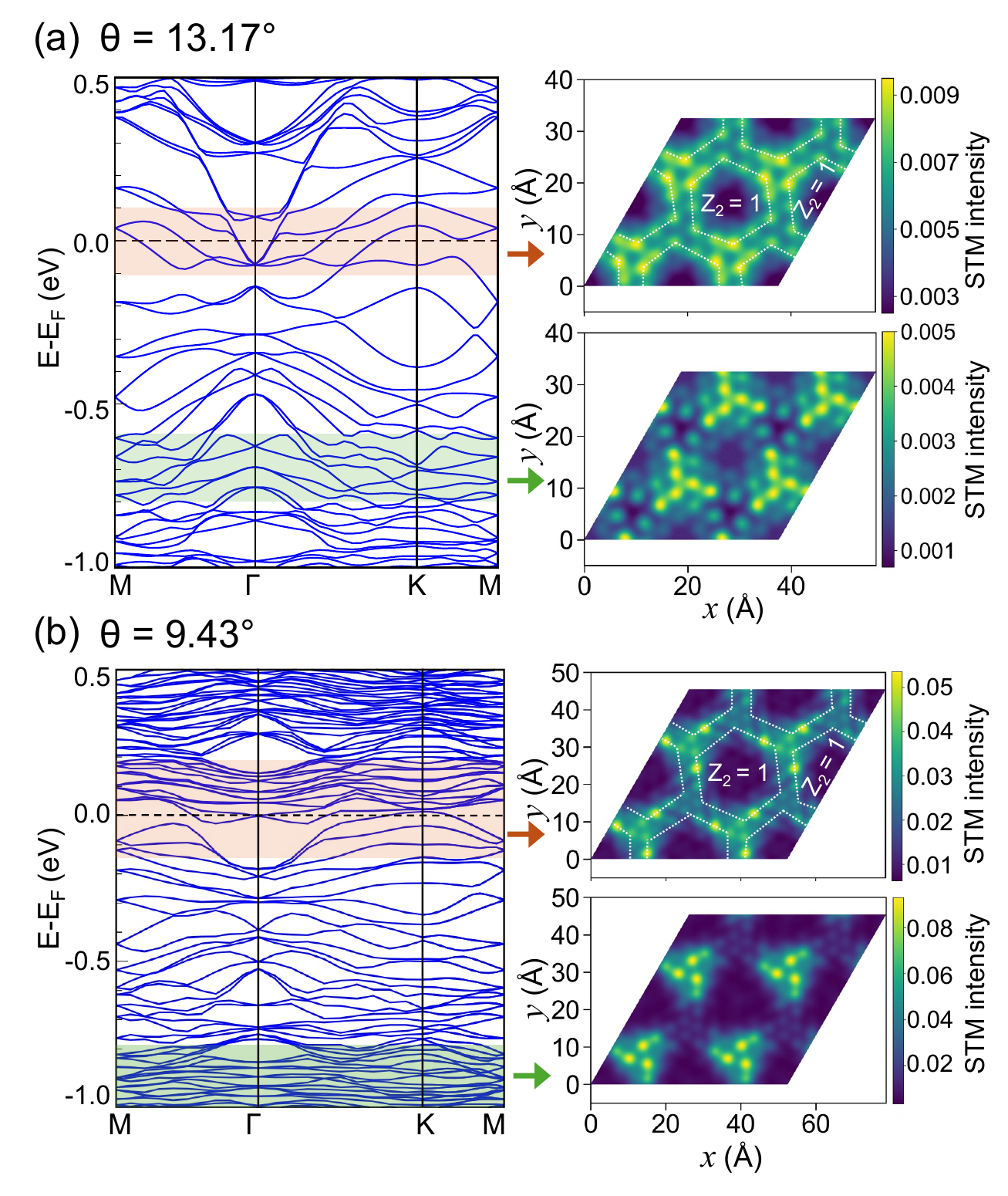}
\caption{Electronic band structures with SOC for twisted BiSb–SbBi bilayers at (a)\,13.17$\degree$ and (b)\,9.43$\degree$. The right panel of (a) shows simulated STM images for a 2\,$\times$\,2\,$\times$\,1 supercell of the 13.17$\degree$ twisted BiSb–SbBi bilayer, probing electronic states near the Fermi level (E$_F\,\pm $\,0.1 eV, red highlighted region) and deeper occupied states (E$_F$\,-\,0.6\,eV to E$_F$\,-\,0.8\,eV, green highlighted region) at a tip height of 4.5 Å. The right panel of (b) shows STM images for the 9.43$\degree$ case probing states near the Fermi level (E$_F\,\pm $\,0.15 eV) and deeper occupied states (E$_F$\,-\,0.8\,eV to E$_F$\,-\,1.0\,eV). Edge states are indicated by white dotted lines. 
Bright yellow dots mark the position of Bi atoms in the top layer, located closest to the STM tip. }

%\caption{(a) and (d) represents the band structure with SOC for \,13.17$\degree$ and 9.43$\degree$ respectively. STM images for 2\,$\times$2\,$\times$1 supercells of twisted BiSb–SbBi bilayers simulated at a tip height of 4.5\,Å. (b)\,13.17$\degree$ twist angle, probed near the Fermi level (E$_F\,\pm $\,0.1 eV) (highlighted in red in the band structure). (c)\,13.17$\degree$ twist angle, probed further below E$_F$ (E$_F$\,-\,0.7\,eV to E$_F$\,-\,0.3\,eV)(highlighted in green in the band structure), (e)\,9.43$\degree$ twist angle, probed near the E$_F$ (E$_F\,\pm $\,0.15 eV). (f)\,9.43$\degree$ twist angle, probed below E$_F$ (E$_F$\,-\,0.4\,eV to E$_F$\,-\,0.3\,eV).}
\label{fig:STM}
\end{figure}

\noindent\textbf{Real space mapping of local edge states in the moir\'e supercell.}
To explore an experimentally feasible method for probing the spatially distributed topological domains in the moir\'e supercell, we simulated STM images for all three twist angles. The results for 13.17$\degree$ and 9.43$\degree$ (Fig.~\ref{fig:STM}) are shown in the main text, whereas the 21.78$\degree$ case is shown in the Supplementary Information. Fig.~\ref{fig:STM}(a) shows the simulated STM images of the twisted bilayer at 13.17$\degree$, probed at different chemical potentials. 
%Panel (a) corresponds to probing electronic states near E$_F$ (E$_F\,\pm $\,0.1 eV), whereas panel (b) corresponds to states further below E$_F$ (E$_F$\,-\,0.7\,eV to E$_F$\,-\,0.3\,eV).

For the 13.17$\degree$ twist angle, probing near the Fermi level reveals highly local density of states (bright yellow) along the boundaries of the Z$_2$\,=\,1 domains at their interfaces with Z$_2$\,=\,0 regions, corresponding to edge states localized at the boundaries of topologically nontrivial domains. In contrast, probing further below E$_F$ excludes these edge states, leaving only bulk trivial states, so no high-LDOS features appear along the boundaries of the nontrivial regions. This establishes STM as a direct probe of emergent edge states in moir\'e-induced topological domains. 
%The bright yellow regions in panel (b) primarily reflect Bi atoms in the top layer near the STM tip. 
A similar behavior is observed for the 9.43$\degree$ twist angle. Enhanced LDOS near E$_F$ (E$_F\,\pm $\,0.15 eV) traces the domain-wall edge states, while at lower energies (E$_F$\,-\,0.8\,eV to E$_F$\,-\,1.0\,eV) the edge-state signature vanishes due to the dominance of bulk states. However, at the larger twist angle of 60$\degree$ corresponding to Bernal stacking, no such edge-state-induced LDOS enhancement is observed, consistent with its topologically trivial character (see Fig.\,S2 in the Supplementary Information). \\
%Overall, the STM images near the Fermi level closely reproduce the spatial patterns observed in the Z$_2$ phase maps.\\
%likely due to interactions between edge states of neighboring Z$_2$\,=\,1 domains. Strikingly, probing further below E$_F$ (E$_F$\,-\,0.4\,eV to E$_F$\,-\,0.3\,eV) reveals more localized edge states.
%produces a uniform STM map corresponding solely to bulk trivial states. 

\begin{figure}[!t]
\centering
\includegraphics [width=1\columnwidth]{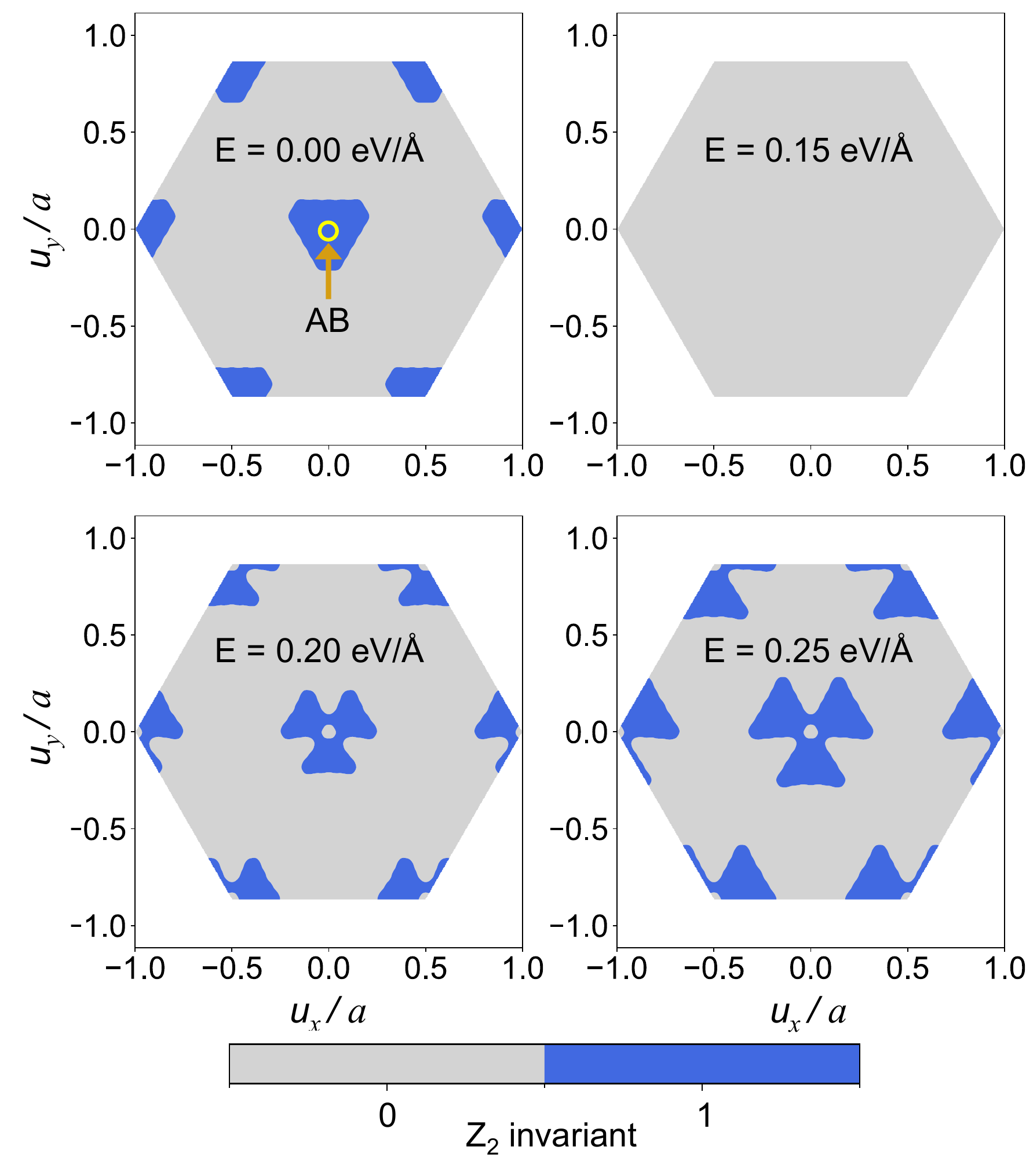}
\caption{Variation of the topological character of different stacking configurations under an external out-of-plane electric field. The centre of the hexagon corresponds to AB stacking, and other points represent randomly shifted configurations relative to AB. \textit{x} and \textit{y} axes denote shifts along the $a_1$ and $a_2$ lattice vectors (fractional units). Blue regions indicate Z$_2$\,=\,1, while gray regions indicate Z$_2$\,=\,0.}
\label{fig:Efield}
\end{figure}

\noindent\textbf{Tunability of topological character through external electric field.}
An external electric field can serve as a tunable knob to control the topological character of the moir\'e supercell. To investigate this effect, a vertical displacement field was applied individually to each of the shifted configurations. Since the moir\'e supercell is effectively a collection of different local stacking configurations, analyzing the electric-field response of the shifted configurations provides an intuitive understanding of the tunability of the topological phase in twisted systems. Fig.~\ref{fig:Efield} shows the variation of the topological character of different stacking configurations under an applied electric field. 
%The center of the hexagonal plot corresponds to the AB stacking configuration, while all other points represent configurations obtained by randomly shifting the two layers relative to AB stacking.

As shown in Fig.~\ref{fig:Efield}(a), at zero field, the AB domain is topologically nontrivial, while the remaining stacking patterns exhibit a mixture of trivial and nontrivial domains. Consequently, the moir\'e supercell hosts both Z$_2$\,=\,1 and Z$_2$\,=\,0 domains in the absence of an applied field. Increasing the field strength to approximately 0.10–0.15 eV/Å renders all stacking configurations trivial, corresponding to a complete OFF state of the conducting edge channels in the moir\'e supercell. Upon further increasing the field, nontrivial domains reemerge, effectively turning the edge channels back to the ON state. 

It is worth noting that the electric field plays a dual role here: (i) it enhances the out-of-plane interlayer tunneling, driving the topological phase transition~\cite{Bordoloi_2DMaterials_2025}, and (ii) it shifts the bands near the Fermi level that are responsible for topological band inversion. As these bands are pushed away from the Fermi level, the band inversion vanishes and the system transitions into a topologically trivial phase. \\ \\
% This demonstrates that the conducting edge channels in the moir\'e supercell can be reversibly tuned by applying external gate voltage.\\

%\noindent\textbf{Topological networks through relaxation.}

\noindent\textbf{Continuum modelling of emergent topological networks and real-space imaging of gapless edge states.}
The complex nature of the multi-domain textures
revealed in Figs.~\ref{fig:phase diagram}-\ref{fig:STM} raises the question of the underlying topological network structure in moiré matter. 
Given the numerical difficulty in accurately determining the topological properties from a full ab-initio calculation, here we provide a conceptual model that captures all the essential physics emerging from full DFT calculations.
We propose a minimal continuum model reflecting the strong SOC and the natural composition of 
%Sb-$p_z$ 
Bi-$p_z$ orbitals in the low energy sector of the (untwisted) system \cite{Bordoloi_2DMaterials_2025}. 
Writing $\vsig=(\sigma_x,
\sigma_y,\sigma_z)$ for the orbital pseudospin Pauli matrices generated by the Bi-$p_z$ orbitals localized on individual layers \cite{Bordoloi_2DMaterials_2025}, 
we construct a spatially modulated 2D analogue of the Bernevig-Hughes-Zhang (BHZ) model (variant of the Kane-Mele model) on the continuum, \cite{bernevig2006quantum,Kane2005}:
\begin{equation}
  H(\mathbf{r}) \;=\; \hv\;\vsig\!\cdot\!\vk\,\hat{\mathbf{1}}\;+\;M(\mathbf{r})\,\sigma_z,
  \quad \vk=-i\nabla.
  \label{eq:H}
\end{equation}
The local $\Ztwo$ of the gapped phase is set by $\mathrm{sgn}\,M$:
$M>0$ trivial, $M<0$ inverted, by convention.

Across an isolated $M=0$ contour, a single
chiral Dirac mode appears in a single time-reversal fixed sector, while its partner adds the
counter-propagating mode in the second sector. Together, they form
a helical Kramers pair localized within a characteristic length scale 
\begin{equation}
  \xi(\mathbf{r}) \;\simeq\; \frac{\hv}{|M(\mathbf{r})|}
  \label{eq:xi}
\end{equation}
at the domain wall~\cite{jackiw1976solitons}. 
The observation of a sharp edge-state network implies $\xi\!\ll\!a_M$, where $a_M$
is the moir\'e periodicity. 
By construction in  Fig.~\ref{fig:STM}, $a_M \approx 2$~nm. 
Using $\hv=120$~meV$\cdot$nm along with a typical mass scale $|M|\sim 1$~eV, which is consistent with the strength of interorbital repulsion in this system \cite{Bordoloi_2DMaterials_2025}, 
we obtain $\xi\!\sim\!0.1$\,nm, placing the system well within the sharp edge-state network regime. 

We parameterize $M(\mathbf{r})$ using first-shell harmonics on the triangular moir\'e 
lattice with reciprocal vectors
$\vb_1,\vb_2$, and $\vb_3=-(\vb_1+\vb_2)$:
\begin{equation}
  M(\mathbf{r}) \;=\; M_0 \;+\; V_{\!c}\sum_{i=1}^{3}\cos(\vb_i\!\cdot\!\mathbf{r})
                     \;+\; V_{\!s}\sum_{i=1}^{3}\sin(\vb_i\!\cdot\!\mathbf{r}).
  \label{eq:Mansatz}
\end{equation}
%
%This is the most general first-shell, $C_3$-symmetric scalar field on the triangular lattice. 
This expression represents the most general first-shell scalar field preserving $C_3$ symmetry on a triangular lattice.
We note that $V_c$ is generically present for any moir\'e lattice with a triangular unit cell. The chiral (and inversion symmetry breaking) properties of the twisting are encoded in $V_s$. 
To leading order, $V_s/V_c \propto \theta$.

Structural relaxation, captured by our first-principles calculations, strongly renormalizes the values of $V_c$ and $V_s$, and naturally incorporates the mirror-symmetry breaking between different stacking regions. 
%particularly in breaking any mirror symmetry relation between them. 
As a result, the AB and BA domains become inequivalent and acquire distinct local $Z_2$ character.
At the same time, the smooth spatial variation of $M(\mathbf{r})$ enables multiple helical edge channels to merge at the boundaries of topologically trivial and nontrivial domains. 
Within this framework of $M(\mathbf{r})$, $M_0$ preserves full rotational symmetry, $V_c$ retains $C_{6v}$ symmetry, and $V_s$ lowers the point group to $C_{3v}$. 
The competition between the relative magnitude of these terms governs both the symmetry breaking induced by twisting and the resulting topological networks in moir\'e matter. 

\begin{figure}[!t]
\centering
\includegraphics [width=1\columnwidth]{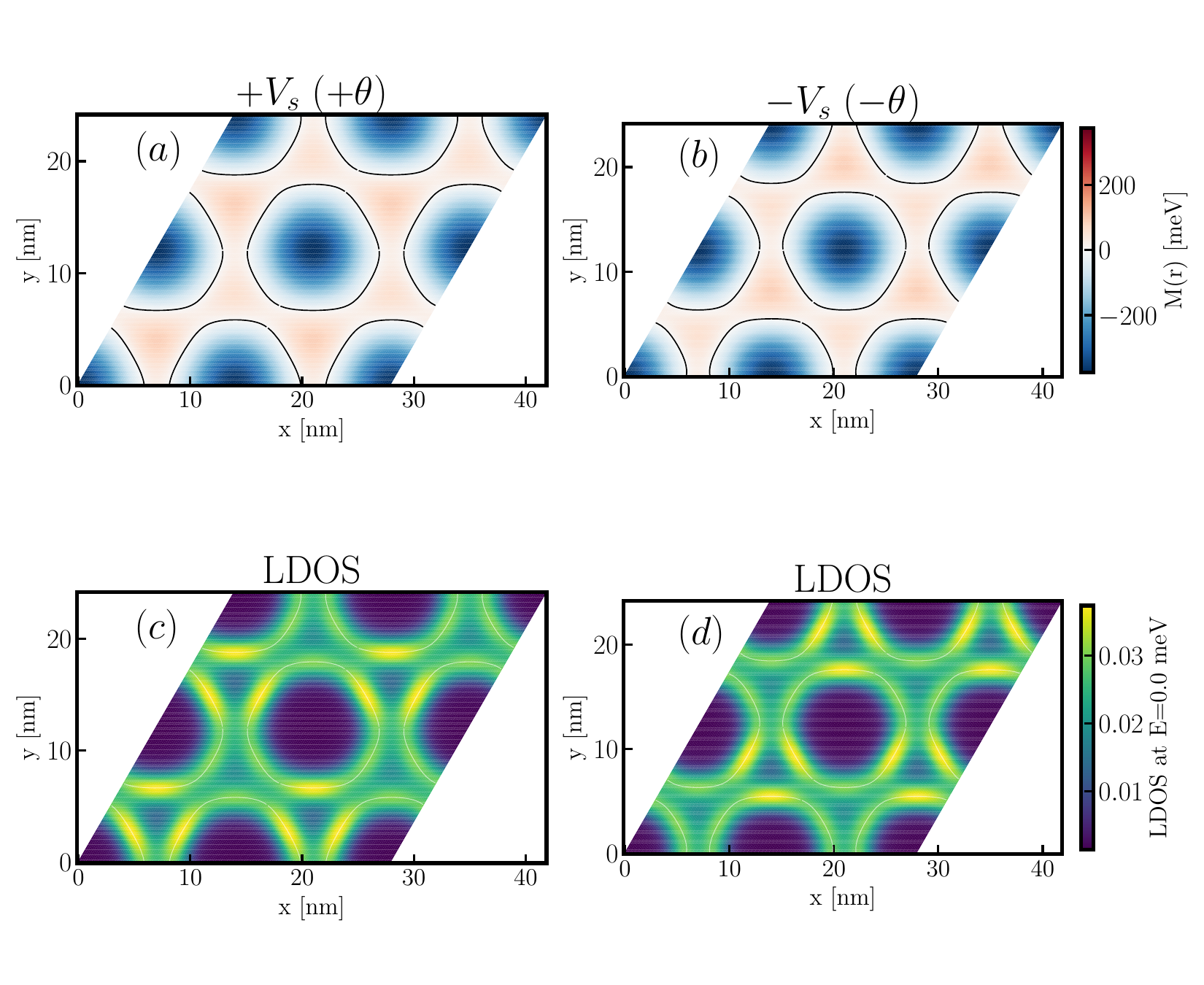}
\caption{Mass distribution $M(\mathbf{r})$ and LDOS for the effective model of Eq.~\eqref{eq:H} which reproduce the full DFT calculations. Parameters are: $V_c =-100$, $V_s = \pm6$, \text{and} $M_0 = -75$ (all meV). For visualization purposes we take $a_M = 14\,\textrm{nm}$. $\hbar v=120\,\textrm{meV nm}$. 
(a-b) Mass distribution as a function of position showing a clear inversion (blue regions) and trivial domain walls (faint red). (a) and (b) show $+\theta$ and $-\theta$ twists, respectively. (c-d) Corresponding LDOS maps showing the inversion-odd partners. Bright yellow spots indicate the topological edge states found near $E=0$. }
\label{fig:bhz}
\end{figure}

We now proceed to solve Eq.~\eqref{eq:H} 
by diagonalizing the Hamiltonian in the moir\'e basis. Letting the wavefunction be $\psi_{\vk,n}(\mathbf{r})
  \;=\; \sum_{\vG}\sum_{l=\uparrow,\downarrow}
        c^{\,\vk,n}_{\vG,\sigma}\,
        e^{i(\vk+\vG)\cdot\mathbf{r}}\,\Ket{l},$ where $l$ is the layer basis encoded by $\boldsymbol{\sigma}$. 
The Hamiltonian becomes: \\
\begin{align}
     \notag &\big[H(\vk)\big]_{(\vG l),(\vG',l')}
  \;=\; \\ &\delta_{\vG\vG'}\,\hv\,\bigl[(\vk+\vG)\cdot\vsig\bigr]_{ll'}
  \;+\; M_{\vG-\vG'}\;[\sigma_z]_{ll'},
  \label{eq:resolved_H}
\end{align}
and the LDOS is: \\
$   \rho(\mathbf{r},E)
  \;=\; \frac{1}{N_k}\sum_{\vk}\sum_{n}|\psi_{\vk,n}(\mathbf{r})|^2\,
        \delta\!\left(E-E_{\vk,n}\right)$. \\

In Fig.~\ref{fig:bhz}, we plot the diagonalization of Eq.~\eqref{eq:resolved_H} for particular parameters chosen to reproduce the DFT results of Fig.~\ref{fig:STM} (see caption). We find an expected inverted mass distribution and a honeycomb-like pattern reflecting the underlying $C_{3v}$ symmetry of the moiré lattice. 
This mass distribution, contains a component which flips sign with reversing the twist direction, directly revealing its chiral origin [Figs.~\ref{fig:bhz}(a)-(b)]. 
We then compute the LDOS, near $E\approx 0$, and plot it in Figs.~\ref{fig:bhz}(c)-(d). 
We observe pronounced one-dimensional edge states which emerge precisely along the boundaries separating topologically nontrivial and trivial domains, whereas the domains themselves remain gapped and exhibit zero DOS (purple hue). 
The observed asymmetry between inequivalent sites originates from the relative magnitude of $V_s$ and $V_c$, which produces a slightly larger positive (trivial) mass in regions where the edge states percolate. This behaviour is consistent with a network-scattering picture in which edge modes preferentially localize along boundaries with the strongest mass inversion, reminiscent of the Chalker-Coddington model~\cite{chalker1988percolation}. 
Importantly, the LDOS maps display a clear contrast under reversal of the twist chirality, providing a direct real-space probe of the \textit{local} moiré chirality.

Taken together, these results demonstrate that scanning probe measurements can be used to reconstruct the spatial mass profile $M(\mathbf{r})$ from LDOS data, similar to recent proposals for mapping electrostatic potentials in moir\'e systems~\cite{seewald2025mapping,klein2026imaging}.

Having established the real-space topological structure, we next analyze the topology of the moir\'e bands in momentum space. 
Within the continuum framework, the evaluation of the $Z_2$ invariant in the full Brillouin zone becomes numerically tractable. 
Supplementary Fig.\,S6 presents the moir\'e band structure together with the corresponding Chern numbers and the $Z_2$ invariants for states filled
up to a particular Fermi level. 
We identify an extended energy window with topological bands {\it below} the charge neutrality point, spanning an energy range of approximately $V_c$ below $E_F$.
We then show the existence of a broad trivial sector of occupied bands.
These results match the observed coexistence of topological and trivial regions identified in DFT calculations shown in Fig.~\ref{fig:STM}(b). 
Combined with gate-tunable STM measurements, our framework provides an innovative route to reconstructing the topological landscape of moir\'e bands and extracting the underlying parameters governing the spatial mass texture $M(\mathbf{r})$. \\

\noindent \textbf{Discussion.} 
Our results establish structural relaxation as a fundamental mechanism that reshapes topology in twisted vdW heterostructures. Rather than producing a uniform topological phase, the interplay of moiré reconstruction, interlayer hybridization, and spin-orbit coupling generates a spatially textured topological landscape in which topologically nontrivial and trivial regions coexist within a single moiré unit cell. 
This establishes a ``topological mosaic” in which the local stacking environment directly controls the sign and magnitude of an effective mass term $M(\mathbf{r})$, making topology intrinsically real-space dependent.
In particular, the spatial modulation of $M(\mathbf{r})$ demonstrates how local stacking environments can act as nanoscale switches between distinct topological regimes, emphasizing the need for real-space approaches when characterizing topology in moiré systems.

%This spatial structure has direct experimental consequences. Domain boundaries host conducting edge states, providing a clear STM fingerprint of the underlying topological texture and enabling reconstruction of $M(\mathbf{r})$ from real-space STM imaging. The sensitivity of the domain network to twist angle and electric field further highlights a route to external control, where topology and edge connectivity can be continuously tuned and electrically reconfigured.
 
A caveat that is worth noting in pristine BiSb bilayers is that these are semimetallic in nature, with some topologically trivial electron/hole pockets lying near the Fermi level. However, as we demonstrate in the Supplementary Figure\,S5,
this semimetallicity can be systematically reduced through chemical substitution with more electronegative elements such as iodine or fluorine, which open a band gap and drive the system toward a semiconducting regime. 
More broadly, the coexistence of nontrivial topology with metallic or lightly doped regimes is of significant interest, as it provides a potential platform for emergent correlated phases, including topological superconductivity~\cite{guerci2024topological}, ferroelectricity~\cite{zhang2024polarization}, and magnetism~\cite{peng2026itinerant}. \\

\noindent \textbf{Conclusions.} 
We have demonstrated that twisting and structural relaxation in BiSb bilayers provide a route to programmable topological matter in moiré systems. 
Starting from two individually trivial monolayers ($Z_2=0$), inversion-stacked and twisted bilayers develop a spatially varying topological phase landscape characterized by coexisting topologically nontrivial ($Z_2=1$) and trivial ($Z_2=0$) domains. 
Relaxation-driven modulation of interlayer separation plays a central role in stabilizing this phase separation in moir\'e matter.
%and in generating internal interfaces that host robust conducting edge states. 

Our continuum BHZ-type description, informed by first-principles calculations, reproduces the real-space topology and establishes a direct connection between STM observables and the underlying spatially varying mass term $M(\mathbf{r})$. This will further extend recent methods for the determination of topology through scanning proble measurements \cite{holbrook2026real,cualuguaru2026observation}.  
The ability to tune domain size via twist angle and reversibly control edge-state connectivity using an electric field demonstrates a high degree of external programmability. 
More broadly, our results identify relaxation as a powerful design principle for engineering reconfigurable topological networks in vdW materials, opening pathways toward intrinsically patterned topological circuits and planar device architectures.

\section{Methods}
\label{comp details}

First-principles density functional theory (DFT) calculations were performed for the moir\'e supercells corresponding to three twist angles:\,21.78$\degree$, 13.17$\degree$, and 9.43$\degree$ using the projector augmented-wave (PAW)  method, as implemented in the Vienna Ab initio Simulation Package (VASP) \cite{Kresse96a, Kresse96b}. The kinetic energy cutoff for the plane-wave basis set was set to 650 eV. The electronic energy convergence criterion was 10$^{-7}$ eV, and the Hellmann-Feynman residual forces on each atom were converged to 10$^{-3}$ eV/Å. 
For structural relaxations of the moir\'e supercell, a $\Gamma$-centered 1\,$\times$\,1\,$\times$\,1 $k$-mesh was used, while a denser 8\,$\times$\,8\,$\times$\,1 mesh was employed for partial charge density calculations. To study the stacking patterns corresponding to different atomic registries (i.e., AA, AB, etc.), a 12\,$\times$\,12\,$\times$\,1 $k$-mesh was used. The exchange–correlation potential was treated within the generalized gradient approximation (GGA) as parameterized by Perdew, Burke, and Ernzerhof for solids (PBEsol)~\cite{PBEsol}. van der Waals (vdW) interactions were included using the DFT-D3~\cite{Grimme_D3} approximation. A vacuum spacing of 15 Å was maintained along the out-of-plane direction. The PAW pseudopotentials included contributions from five valence electrons for both Bi ($6s^26p^3$) and Sb ($5s^25p^3$) atoms. VASPKIT~\cite{VASPKIT} was used to simulate STM images by post-processing the partial charge density data.

Wannierization of the individual stacking configurations was performed using maximally localized Wannier functions, as implemented in Wannier90~\cite{Wannier90}, with initial projections taken from the s, $p_x$, $p_y$, and $p_z$ orbitals of Bi and Sb. The Z$_2$ topological invariants were subsequently computed from the Wannier tight-binding Hamiltonian using WannierTools~\cite{WU2018405}. To study the effect of an external vertical electric field, a displacement field was applied to the Wannier tight-binding Hamiltonian,
\begin{align}
H_{\alpha \beta}(\mathbf{k}) = \sum_{\mathbf{R}} t_{\alpha \beta}(\mathbf{R})e^{i\mathbf{k}\cdot\mathbf{R}},
\label{eq:wan_ham}
\end{align}
where $R_i = na_1 +ma_2$ denotes lattice vectors, $t_{\alpha \beta}$ are hopping amplitudes in the Wannier basis, and $\alpha \beta$ represent orbital indices. Assuming the Wannier orbitals are well localized along the $z$ axis, the effect of the vertical displacement field $E$ was introduced as
\begin{align}
H^{\alpha \beta}(\mathbf{k}) = H_0^{\alpha \beta}(\mathbf{k}) + \delta_{\alpha\beta} e R^z_{\alpha} E,
\end{align}
where $R^z_{\alpha}$ is the $z$ coordinate of the $\alpha$ Wannier orbital and $H_0$ is the unperturbed Hamiltonian in Eq.~\eqref{eq:wan_ham}.

\section{Acknowledgements}
%\noindent\textbf{Acknowledgements}
A.B.~and S.S.~acknowledge support from the U.S.~Department of Energy, Office of Science, Office of Fusion Energy Sciences, Quantum Information Science program under Award No.~DE-SC-0020340. Authors also acknowledge support from the Furth Research Fund at the University of Rochester. 
D.K.~is supported by an Abrahams postdoctoral fellowship of the Center for Materials Theory, Rutgers University and the Zuckerman STEM fellowship. D.K.~thanks the hospitality of the Aspen Center for Physics, which is supported by National Science Foundation grant PHY-2210452, where a part of this work was carried out.
A.B.~and S.S.~thank the Pittsburgh Supercomputer Center (Bridges2) supported by the Advanced Cyberinfrastructure Coordination Ecosystem: Services \& Support (ACCESS) program, which is supported by National Science Foundation grants \#2138259, \#2138286, \#2138307, \#2137603, and \#2138296.  

 \bibliography{bibfile}% Produces the bibliography via BibTeX.
\end{document}